# MAVKA: software for statistically optimal determination of extrema

Kateryna D. Andrych[1,2], Ivan L. Andronov[2]

(1) Department of Theoretical Physics and Astronomy, Odessa I.I.Mechnikov National University, 2 Dvoryanskaya St., Odessa 65026, Ukraine, katyaandrich@gmail.com

(2) Department "Mathematics, Physics and Astronomy", Odessa National Maritime University, Mechnikova, 34, 65029 Odessa, Ukraine, tt_ari@ukr.net

**Abstract:** We introduce the program MAVKA for determination of characteristics of extrema using observations in adjacent data intervals, with intended applications to variable stars, but it may be used for signals of arbitrary nature. We have used a dozen of basic functions, some of them use the interval near extremum without splitting the interval (algebraic polynomial in general form, "Symmetrical" algebraic polynomial using only even degrees of time (phase) deviation from the position of symmetry argument), others split the interval into 2 subintervals (a Taylor series of the "New Algol Variable", "the function of Prof. Z. Mikulášek"), or even 3 parts ("Asymptotic Parabola", "Wall-Supported Parabola", "Wall-Supported Line", "Wall-Supported Asymptotic Parabola", "Parabolic Spline of defect 1"). The variety of methods allows to choose the "best" (statistically optimal) approximation for a given data sample. As the criterion, we use the accuracy of determination of the extremum. For all parameters, the statistical errors are determined. The methods are illustrated by applications to observations of pulsating and eclipsing variable stars, as well as to the exoplanet transits. They are used for the international campaigns "Inter-Longitude Astronomy", "Virtual Observatory" and "AstroInformatics". The program may be used for studies of individual objects, also using ground-based (NSVS, ASAS, WASP, CRTS et al.) and space (GAIA, KEPLER, HIPPARCOS/TYCHO, WISE et al.) surveys.

## Introduction

Variable stars of different types are the most important sources of information about structure of stars and their evolution. Thus they are extensely observed in many countries.

For some types, like eclipsing or pulsating variables, there are many short time series of observations, which do not cover the whole photometrical period of the object. Typically this is due to a separation of the tasks to analyse precise multi-colour observations to determine parameters and get only one parameter from the incomplete light curve the "Time of Minimum/Maximum" (ToM) (cf. Tsesevich, 1970, 1971). The moments of extrema (also referred as "minima/maxima timings") are used for the "O-C" studies of the period changes due to mass transfer and/or presence of a third/fourth body. The largest compilations were published as a paper monograph by Kreiner et al. (2001), whereas further catalogs are regularly improved and are accessible on-line. Among the most famous are the "O-C gate" (project BRNO, Czech Astronomical Society & Masaryk University) with an on-line possibility to determine the ToM (Brát et al., 2012), and the AAVSO (2018) database with recommended external software.

The popular methods of determination of ToM varied from "bisector" (Pogson's) method using the light curve and its hand-written approximation to a polynomial (typically parabolic) approximation (e.g. Tsesevich 1970, 1971) or the method by Kwee and van der Woerden (1956). The software for determination of ToM using the polynomial fit, was presented e.g. by Breus (2006). The statistically optimal degree of the polynomial was used to compile a catalogue of characteristics of 173 semi-regular pulsating stars (Chinarova and Andronov, 2000). A review on long-period variables was presented by Andronov et al. (2014). The improvement of the "bisector" method (using only times of crossing of some constant value) was proposed by Andronov and Andrych (2014).

Newer methods of the ToM determination are based on the least squares, and were reviewed by Andronov (1994, 2005). For the complete light curves of intermediate polars, two-period (multi-)harmonic approximations were used (e.g. Andronov and Breus, 2013). For the complete light curves of the eclipsing binaries, Andronov (2010, 2012) proposed the "New Algol Variable" (NAV) algorithm, which is effective not only for Algols, but also for EB and EW-type systems (Tkachenko et al., 2016). Andronov et al. (2017) tested many functions to make a statistically optimal phenomenological approximation of the complete eclipse.

We focused on approximations for a common case, when time series are close to the extremum of brightness of the object. From such data needed to determine characteristics of extremum: moment, magnitude, width of flat part of eclipse (if there is transit) and corresponding accuracy estimates. Moreover, some types of functions are asymmetrical, what is generally the case for the pulsating variables, and may also be applied for eclipses of the binaries with the O'Connell effect.

For this task, the program MAVKA ("Multi-Analysis of Variables by Kateryna Andrych") was elaborated initially in the Excel/VBA environment (Andrych et al., 2015) and then was rewritten in the Delphi 7 environment (Andrych et al., 2017) with adding these approximations to the Delphi version of the "Observation Obscurer" (OO) software, which was originally written in Free Pascal (Andronov, 2001). The software MAVKA allows to compute necessary parameters of the extrema using one out of nine realized methods.

In this paper, we discuss this program and algorithms, realised in it.

**Methods of the analysis**

Nine different methods were realized in program.

For all approximations linear parameters are determined by the method of least squares, snd for non-linear - by the method of differential corrections.

- (1) Algebraic polynomial approximation of different degrees α. The program automatically chooses the most precise degree, which corresponds to the best accuracy of the moment of extremum.

$$x_c(u) = \sum C_\alpha \cdot f_\alpha(u), \; f_\alpha(u) = u^{\alpha-1}$$

- (2) "Symmetrical" algebraic polynomial approximation, which use only even degrees of deviation from point of symmetry.

$$x_c(u) = \sum C_\alpha \cdot f_\alpha(u), f_\alpha(u) = (u - u0)^{2(\alpha-1)}$$

The next two methods are

- New Algol Variable, that was proposed by Andronov (2010, 2012):

$$x_c(u) = C_1 + C_2\left(1 - \left(1 - \left(\frac{u - C_3}{C_4}\right)C_5\right)^{1.5}\right)$$

- the function proposed by Mikulášek (2015), Mikulášek et al. (2015):

$$x_c(u) = C_1 + C_2\left(1 - \left(1 - \exp\left(1 - \text{ch}\left(\frac{u - C_3}{C_4}\right)\right)\right)C_5\right)$$

In these forms, the functions are applied for approximation of the whole light curve. In the case, where the data are just close to the extremum, the matrix of normal equations becomes degenerate. The interval does not contain out-of-eclipse parts, and therefore we do not have enough data. Formally, the amplitude becomes infinite, and the error estimates are not available because of rounding errors.

To avoid degeneration of the matrix of the normal equation, the last two functions were expanded into the Taylor series, and the following approximations are used:

- (3) New Algol Variable (Andronov et al., 2017)

$$x_c(u) = C_1 + C_2|z|^{C_6} + C_3|z|^{2C_6}, \quad z = u - C_5$$

- (4) the function proposed by Prof. Z. Mikulášek (2015)

$$x_c(u) = C_1 + C_2|z|^{C_6} + C_3|z|^{C_6+2}, \quad z = u - C_5$$

These forms of functions realized in MAVKA.

In some cases, for example approximation of flat eclipses and exoplanet transitions, it is more precise to divide our data into three intervals, each of them is approximated by its own function. In this case, an additional parameter is the duration of the bottom part of the eclipse (total eclipse or a transit), This parameter is suggested to be listed in the "General Catalogue of Variable Stars" (GCVS, Samus' et al., 2017).

The additional parameters (the borders between the three intervals) are determined by minimizing the test function (sum of the squares of the residuals of the data from the approximation) in respect to these parameters, and then corrected using differential equations. For the visualization, we have applied an own "gradient zebra" – type colour scheme (Andrych et al., 2017) instead of a classical "lines of constant level" (e.g. Tkachenko, 2016).

We have proposed some new methods for this case.

- (5) The Wall-Supported Parabola method (Andrych et al., 2017): part of data that is very close to extremum is approximated by parabola, but branches are approximated by a sum of a parabola and a contribution of another degree.

    Let $C_5$ and $C_6$ be transition points from one function to another.

    In the case of transition or eclipse for a small deviation of time from the point of the inner or outer contact, the flux increases/decreases proportionally to the difference $(C_5 - u)$ or $(u - C_6)$ in the non-integer power 1.5.

    If $u < C_5$ then

    $$x_c(u) = C_1 + C_2\left(u - \frac{C_5 + C_6}{2}\right) + C_3(C_5 - u)^{1.5}$$

    If $u > C_6$:

    $$x_c(u) = C_1 + C_2\left(u - \frac{C_5 + C_6}{2}\right) + C_3(u - C_6)^{1.5}$$

    If $C_5 < u < C_6$:

    $$x_c(u) = C_1 + C_2\left(u - \frac{C_5 + C_6}{2}\right)$$

- (6) Wall-Supported Line (Andrych et al., 2017): it is nearly similar, but instead of the parabola, there is a constant line for the central interval. This method is useful for flat minima and it allows one to calculate the duration of the full eclipse or transition if the limb darkening effect can be neglected.

    If $u < C_5$ then

    $$x_c(u) = C_1 + C_2(C_5 - u)^{1.5} + C_3(C_5 - u)^{3.5}$$

    If $u > C_6$:

    $$x_c(u) = C_1 + C_2(u - C_6)^{1.5} + C_3(u - C_6)^{3.5}$$

    If $C_5 < u < C_6$:

    $$x_c(u) = C_1$$

- (7) Wall-Supported Asymptotic Parabola (this paper). The method improves the method of "Asymptotic parabola" (Marsakova and Andronov, 1996) by adding two parameters describing terms at the left and right interval, which correspond to ascending and descending branches. This method may be effective for approximation of the symmetric eclipses. For pulsating variables, it may be computed, but has no physical meaning.

    If $u < C_5$ then

    $$x_c(u) = C_1 + C_2(-D^2 - 2Dv) + C_3(C_5 - u)^{1.5}$$

    If $u > C_6$:

    $$x_c(u) = C_1 + C_2(-D^2 + 2Dv) + C_3(u - C_6)^{1.5}$$

    If $C_5 < u < C_6$:

    $$x_c(u) = C_1 + C_2 v^2$$

    Where $D = \frac{C_6 - C_5}{2}, v = u - \frac{C_6 + C_5}{2}$.

- (8) Parabolic Spline (this paper). This approximation has more parameters than previous ones. Therefore, transition points from one function to another are $C_6$ and $C_7$. The method improves the method of "Asymptotic parabola" (Marsakova and Andronov, 1996) by adding two parameters describing parabolic terms at the left and right interval. This method allows to extend the width of the interval (and so to increase the number of points) to get a same accuracy, as the "Asymptotic parabola".

    If $u < C_6$ then:

$$x_c(u) = C_1 + C_2 u + C_3 u^2 + C_4(C_6 - u)^2$$

If $u > C_7$:

$$x_c(u) = C_1 + C_2 u + C_3 u^2 + C_5(u - C_7)^2$$

If $C_6 < u < C_7$:

$$x_c(u) = C_1 + C_2 u + C_3 u^2$$

- (9) Asymptotic Parabola.

There are two lines "asymptotes", which are connected with a parabola, so the function and it's derivative are continuous. Previous experience had shown, that this is one of the best methods for asymmetrical maxima of generally pulsating variables. This method was originally proposed by Marsakova & Andronov (1996).

The transition points from one interval to another are $C_4$ and $C_5$.

If $u < C_4$ or $u > C_5$ then:

$$x_c(u) = C_1 + C_2(2\,Abs(v) - D)D + C_3 v$$

If $C_4 < u < C_5$ then:

$$x_c(u) = C_1 + C_2 v^2 + C_3 v$$

Where $D = \frac{C_5 - C_4}{2}, v = u - \frac{C_5 + C_4}{2}$.

## MAVKA interface

The program was elaborated in the programming environment Delphi 7.

In MAVKA, the user can choose few methods and the algorithm automatically defines the method, which corresponds to the best accuracy of the moment of extremum (Fig. 1).

In an addition, the user can control results, because while the approximation may be mathematically accurate, formally the values of parameters may be out of physically realistic intervals (Fig. 2).

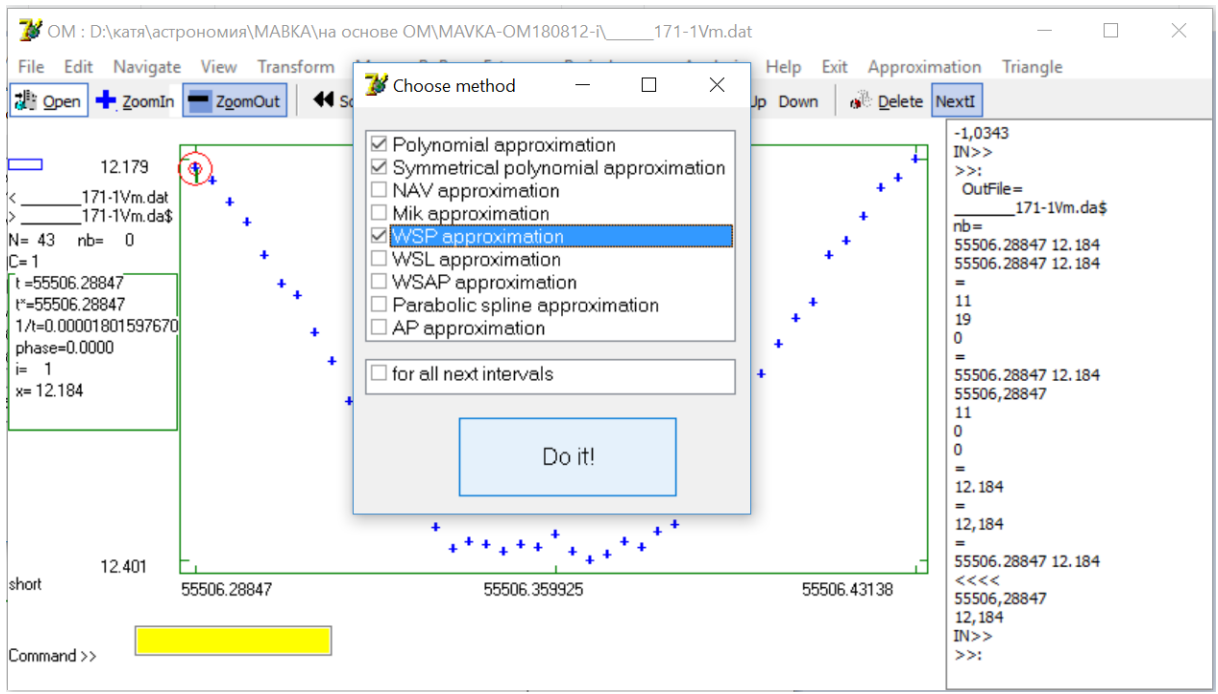

**Figure 1:** MAVKA submenu for choosing some approximation methods.

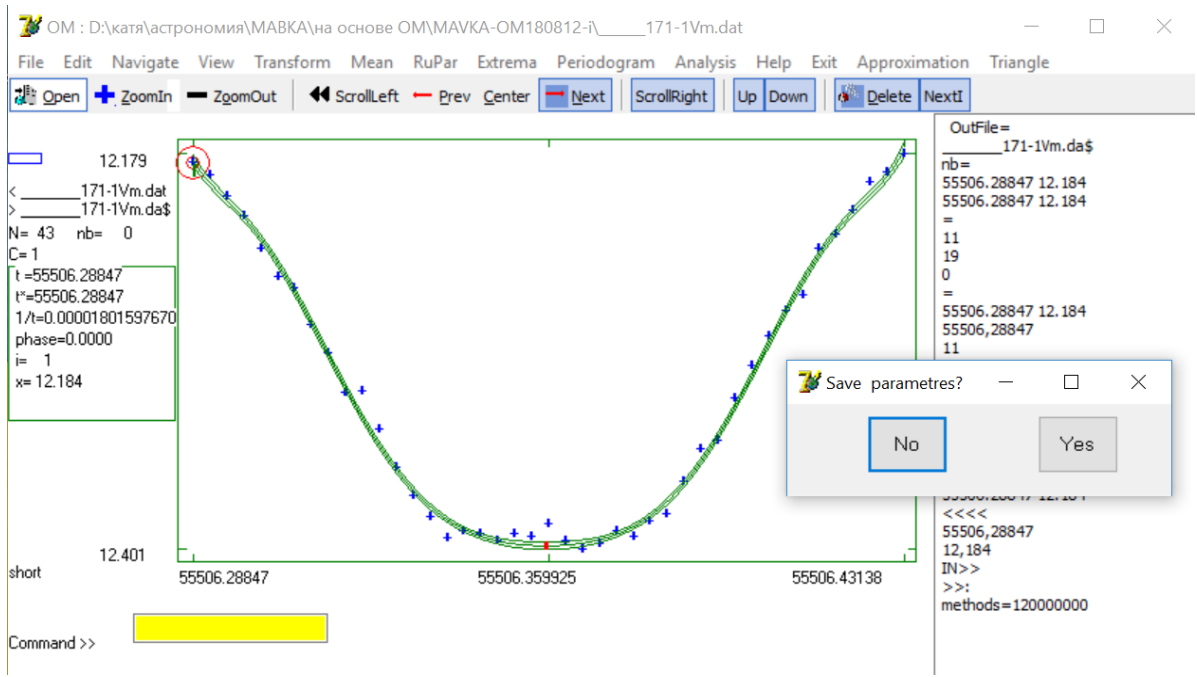

**Figure 2:** The possibility of visual control of approximation and results.

**Application to Variable Stars of Different Types**

Previous versions of the program, which was being improved for difficult cases of the distribution of the times (phases) of the observations in selected intervals near extrema, were applied to observations of some eclipsing binary stars (Savastru et al, 2017; Tvardovskyi et al., 2017, 2018) and symbiotic variables (Marsakova et al., 2015). This new version of the program is planed to be used for compilation of the characteristic of extrema of pulsating, eclipsing and interacting binary stars based on data from photometric surveys and own monitoring.

**Discussion and Conclusion**

We have introduced the program MAVKA. It allows the user to choose statistically optimal phenomenological approximation for a given data sample and determine characteristics of extrema with the best accuracy in case, when the data are just close to the extremum. This is the main mode of the observations by the "hunters" for the "extrema timings" = "Times of Minima/Maxima" = ToMs". Besides the ToM, the brightness at the extremum is determined, and (for the "3-interval" functions (5,6,7,9)) the duration of the full eclipse/transit. The error estimates for each parameter are determined.

The program allows one to determine the parameters either for the data at the screen, or by using a list of previously marked intervals – in this case, one may automatically determine the statistically optimal phenomenological approximations for all intervals and then check the quality of approximation by comparing with the data.

We have presented 9 different approximation functions, but we allow the determination of the statistically optimal degree of the polynomial or "symmetric polynomial", so currently there are 20 functions to be used for the comparison.

The methods (1, 8, 9) may be recommended for approximation of generally asymmetric extrema, whereas the rest (2-7) are focused on the theoretically symmetric extrema.

**Acknowledgement**

We thank Doc. RNDr. Vladyslava I. Marsakova, Lewis M. Cook, Anton Paschke, Mgr. Kateřina Hoňková, Doc. RNDr. M. Zejda and Prof. RNDr. Zdeněk Mikulášek for discussions. This work is related to the international projects "Inter-Longitude Astronomy" (ILA, Andronov et al., 2003, 2017) and "Ukrainian Virtual Observatory", "Astroinformatics" (Vavilova et al. 2017).